\documentstyle[preprint,aps,amsfonts]{revtex}
\def\vec#1{\bbox{#1}}
\def\i{{\mathrm i}}
\begin{document}
\tighten
\title{Current conservation in thermal field theory\thanks{
Work supported by GSI.}}
\author{P.A.Henning\thanks{P.Henning@gsi.de}
  and M.Blasone\thanks{blasone@tpri6c.gsi.de}}
\address{Theoretical Physics,
        Gesellschaft f\"ur Schwerionenforschung GSI\\
        P.O.Box 110552, D-64220 Darmstadt, Germany}
\maketitle
\begin{abstract}
Within the framework of generalized free field theory
at nonzero temperature we address the problem
of current conservation. The formalism of thermo field dynamics is
used to derive a conserved and thermodynamically consistent
physical current operator. Consequences for the calculation
of photon emission rates from a hot plasma are considered briefly.
\end{abstract}
To common wisdom the fundamental interactions of the universe are
described by gauge theories. Most often they are used
perturbatively, i.e., physical quantities of interest are
obtained by expansion in terms of non-interacting stable asymptotic
states. Such a treatment can very successfully account for nearly
all data measured in few-body scattering experiments.
However, gauge theories also play an important role
in physical systems with high temperature and a large number of constituents,
such as the early universe or relativistic heavy-ion collisions.
Unfortunately for the theoretical physicist,
the perturbative approach to gauge theory breaks
down at nonzero temperature \cite{L80}.

Technically this breakdown emerges in the form of severe infrared divergences,
which one can cure to a certain loop order by the introduction
of temperature dependent masses for the gauge bosons.
The method of ``hard thermal loops'' has evolved along these lines.
{}From a more fundamental point of view however this breakdown
is understood through a theorem by Narnhofer, Requardt and
Thirring \cite{NRT83}, which  states that (quasi-)particles
at nonzero temperature must be strictly free of interaction.

Conversely, to obtain a meaningful solution of a gauge theory at
nonzero temperature to all orders, one has to deviate from the picture of
asymptotically stable excitations.  A consequent reformulation of
a scattering theory in terms of such states with an a-priori
spectral width, so-called generalized free fields,
has been achieved by Landsman \cite{L88}. The central quantity of
this method is the spectral function ${\cal A}$, or the imaginary part
of the retarded Green function
$S^R(E,\vec{p}) = \mbox{Re}G(E,\vec{p}) - \i\pi{\cal A}(E,\vec{p})$
in momentum space. Diagram rules, transport theory and other developments
of traditional thermal field theory may be carried over to this
treatment in terms of spectral functions,
and one may also show that this method is indeed non-perturbative in
the sense of not being power-expandable around zero coupling constant
\cite{h94rep}.

However, particular problems arise with such a formulation,
and one of these is addressed with the present paper:
What is the conserved current of an (abelian) gauge theory in
a hot system, where the constituents according to the NRT
theorem do not have a mass shell ? We first solve this problem,
and then construct the physical current operator for thermal field
theory. Finally we examine the consequences of current conservation
for the production rates of gauge bosons (photons).

To this end we consider a generalized free fermion field in the framework of
thermo field dynamics (TFD) for quantum fields with continuous mass spectrum
\cite{TFD,h94rep}. In this formulation,
the field is described by two different representations
$\psi_x$ and $\widetilde{\psi}_x$, having the canonical
anticommutation relations
\begin{equation}
\left\{\psi(t,\vec{x}) , \vphantom{\int}
        \psi^\dagger(t,\vec{x}^\prime)\right\}  =
        \delta^3(\vec{x}-\vec{x}^\prime) \;\;,\;\;\;\;\;
\left\{\widetilde{\psi}(t,\vec{x}) , \vphantom{\int}
\widetilde{\psi}^\dagger(t,\vec{x}^\prime)\right\}  =
        \delta^3(\vec{x}-\vec{x}^\prime)
\;\end{equation}
and anticommuting with each other.  The
statistical doublets of the fermions are,
in slight modification of ref. \cite{h94rep}, introduced as
\begin{equation}\label{doub}
\Psi_x =
\left({\array{c} \psi_x \\ -\i\widetilde{\overline{\psi}}_x\gamma^0
       \endarray}\right)\;\;\;\;\;\;
\overline{\Psi}_x^T =
\left({ \array{c} \overline{\psi}_x \\ -\i\gamma^0\widetilde{\psi}_x
       \endarray } \right)
\;,\end{equation}
where $\psi_x\equiv\psi(x)$.
The two-point function therefore is a 2$\times$2 matrix with
elements
$S^{ab} =
  -\i\left\langle
\mbox{\large T}\left[ \Psi^a_{x}\overline{\Psi}^b_{x^\prime} \right]
 \right\rangle$, where $\left\langle\cdot\right\rangle =
 \langle\mkern-4mu\langle 1|\cdot| W \rangle\mkern-4mu\rangle$
denotes the thermal average (a simple matrix element in the TFD
formalism). This propagator equals the two-point function
of the Schwinger-Keldysh formalism \cite{SKF}, and
by construction obeys $S^{11}+S^{22}-S^{12}-S^{21} = 0$.
Here this identity follows from
\begin{equation} \label{gse}
\langle\mkern-4mu\langle 1|
\left( \psi_x + \i\widetilde{\overline{\psi}}_x\gamma^0\right ) = 0
\;\;,\;\;\;\;
\langle\mkern-4mu\langle 1|
\left( \overline{\psi}_x + \i\gamma^0\widetilde{\psi}_x\right ) = 0
\;.\end{equation}
Since we assume a generalized free field, it is described by
a bilinear lagrangian. In contrast to ordinary free fields however
this lagrangian includes a non-local term,
\begin{equation}\label{lagr}
{\cal L}\left[\Psi\right]=
\overline{\Psi}_x \left( \tau_3 \left( {\mathrm i}\partial_\mu\gamma^\mu
 - m\right)\right) \Psi_x - \int\!\!d^4y\,
\overline{\Psi}_x \,\widehat{\Sigma}(x,y)\, \Psi_y
\;.\end{equation}
In this equation, $\tau_3$=diag(1,-1) and $\widehat{\Sigma}$ is also
a $2\times2$ matrix. Note, that due to the off-diagonal
matrix elements $\Sigma^{12}$ and $\Sigma^{21}$
of the self energy function, tildean and non-tildean
fields are mixed in this lagrangian.

In ${\cal L}$ we now perform a local gauge transformation according to
\begin{equation}
\Psi_x  \rightarrow (1- {\mathrm i} \alpha_x) \Psi_x \;\;,\;\;\;\;
\overline{\Psi}_x  \rightarrow (1+ {\mathrm i} \alpha_x)
 \overline{\Psi}_x
\;,\end{equation}
which leads to a variation of the lagrangian
\begin{eqnarray}\nonumber
\delta {\cal L}_x & =\; \frac{\partial \alpha_x}{\partial x^\mu}&\,
  \left(\vphantom{\int}
\overline{\Psi}_x \tau_3 \gamma^\mu \Psi_x
 + {\mathrm i}\int \!\!d^4y\,\overline{\Psi}_x (y - x)^\mu
  \,\widehat{\Sigma}(x,y)\,\Psi_y \right.\\
&&-\left. \frac{{\mathrm i}}{2}\frac{\partial}{\partial x^\nu}\,
  \int \!\!d^4y\,\overline{\Psi}_x (y - x)^\mu (y - x)^\nu
  \,\widehat{\Sigma}(x,y)\,\Psi_y \right) + {\cal O}\left(\partial^3\right)
\;.\end{eqnarray}
After a Fourier transform (we assume a homogeneous equilibrium system),
we obtain as the conserved current of the generalized free field
\begin{equation}
\label{conc}
J^\mu(x)  =  \overline{\Psi}_x \tau_3 \gamma^\mu \Psi_x
         -  \i\int\!\!d^4yd^4z \frac{d^4p}{(2\pi)^4}\frac{d^4q}{(2\pi)^4}
 \overline{\Psi}_z \widehat{\Lambda}^\mu(p,q)\mbox{e}^{{\mathrm i}p(x-z)
 -{\mathrm i}q(x-y)} \Psi_y
\;.\end{equation}
The function $\widehat{\Lambda}(p,q)$ occuring here has an expansion
in terms of derivatives of the Fourier transformed self energy function
$\widehat{\Sigma}(q)$
\begin{equation}\label{exvex}
\widehat{\Lambda}^\mu(p,q) = \frac{\partial}{\partial q_\mu}
 \widehat{\Sigma}(q) + \frac{1}{2!} (p - q)_\nu
\frac{\partial^2}{\partial q_\mu \partial q_\nu}\widehat{\Sigma}(q)
 + \dots
\;.\end{equation}
Current conservation, i.e., vanishing of $\partial_\mu J^\mu(x)$ on the
operatorial level then amounts to the requirement
\begin{equation}
(p - q)_\mu \widehat{\Lambda}^\mu(p,q)
= \widehat{\Sigma}(p) - \widehat{\Sigma}(q)
\;,\end{equation}
which is the well-known Ward-Takahashi identity for the
matrix valued self energy function. It leads us to the identification
of $\tau_3 \gamma^\mu - \widehat{\Lambda}^\mu(p,q)$
as the irreducible vertex function of thermal field theory.

In principle, this method may be used to derive the conserved current
for any nonlocal theory. However, for the special case of thermal
field theory with the doublet structure of the
lagrangian (\ref{lagr}), it leads to an
unphysical current operator: Its thermal groundstate expectation
value does not depend on the particle density. Note, that
this problem is equally present in other formulations of
thermal field theory \cite{SKF}.

To construct the conserved {\em physical\/} current which is the source of
an external gauge field associated with the local phase
invariance of the lagrangian (\ref{lagr}), we have to split the
unphysical operator (\ref{conc}) in two parts which are time
reversed to each other. The ``half'' with the proper
boundary conditions then may be regarded as the physical current operator
$j^\mu(x)$ of a thermal system with continuous mass spectrum.
This splitting is complicated by the presence of
the mixing terms $\Sigma^{12}$ and $\Sigma^{21}$ in the lagrangian
(\ref{lagr}),

We find, that the following three requirements are sufficient to
determine this operator completely:
\begin{enumerate}
\item
$j^\mu(x)$ is conserved at the operatorial level,
$\partial j^\mu(x)/\partial x^\mu = 0$.\\
\item
$j^\mu(x)$ is a bilinear functional of the fields, and in case of a
diagonal self energy function with $\Sigma^{12} \equiv \Sigma^{21} \equiv 0$
reduces to the $\overline{\psi} \dots \psi$ component of $J^\mu(x)$,
i.e., to ``one half'' of the Noether current. Note, that this requirement
also leads to a reduction to the free physical current
in case of a completely vanishing self energy function,
$\lim_{\Sigma\rightarrow 0} j^\mu(x) = \overline{\psi}_x \gamma^\mu \psi_x$.
\item
$j^\mu(x)$ has a real expectation value in the thermal
ground-state, with a well-defined thermodynamical interpretation.
\end{enumerate}
The operator for the conserved physical current, which follows from
these three requirements is
\begin{eqnarray} \nonumber
j^\mu(x) &=& \overline{\psi}_x \gamma^\mu \psi_x
 - \i\int\!\!d^4y d^4z\, \frac{d^4p}{(2\pi)^4}\frac{d^4q}{(2\pi)^4}\;
 \overline{\psi}_z \widehat{\Lambda}^{11\, \mu}(p,q)\,
\mbox{e}^{{\mathrm i}p(x-z)  -{\mathrm i}q(x-y)} \psi_y \\
\nonumber
&+& \i\delta^{\mu0} \int\!\!d^4y\,\int\!\!d\tau\, \Theta(x_0-\tau) \left[
\overline{\psi}(\tau,\vec{x})\,\vphantom{\int}
\Sigma^{12}(\tau,\vec{x};y) \,
 \left(-\i\widetilde{\overline{\psi}}(y)\gamma^0\right)\right.\\
\nonumber
&&\;\;\;\;\;\;\;\;\;\;\;\;\;\;\;\;\;\;\;\;
\left.- \left(-\i \gamma^0\widetilde{\psi}(y)\right)\,
 \Sigma^{21}(y;\tau,\vec{x})\,
 \psi(\tau,\vec{x})\vphantom{\int}\right]\\
\label{cc0}
&+&  \i\delta^{\mu0} \int\!\!d^4y\,\int\!\!d\tau\, \Theta(y_0-\tau)
 \,\mbox{sign}(y_0)\,\left\{\vphantom{\int}
 \overline{\psi}(\tau,\vec{x}) \Sigma^{12}(\tau,\vec{x};y) , \psi(y) \right\}
\,,\end{eqnarray}
where $\left\{\cdot,\cdot\right\}$ is the anticommutator of the enclosed
quantities, and  $\Theta(\cdot)$ is the Heaviside step function.

The construction of the physical current
according to a {\em temporal\/} boundary condition, as well as the
quantization described here, imply the choice of a certain reference frame.
Most naturally this is
the frame where the three-vector current expectation value
vanishes. This explains the factors $\delta^{\mu0}$ in the above
operator, which however together with the temporal step functions
may be generalized to an arbitrary timelike hypersurface.

The  anticommutator term in $j^\mu(x)$ is,
due to the fact that we consider generalized
free fields, a complex number equal to
\begin{equation}
\overline{j}^\mu = \frac{\delta^{\mu 0}}{\pi}
 \;\int\!\!dE\,\int\!\!\frac{d^3\vec{p}}{(2\pi)^3}
\,\left( \mbox{Re}G(E,\vec{p}) - \i \pi {\cal A}(E,\vec{p})\right)\,
\frac{\partial}{\partial E} \left( n(E) \mbox{Im}\Sigma^R(E,\vec{p}) \right)
\;.\end{equation}
$\mbox{Re}G(E,\vec{p})$ and $ -\pi {\cal A}(E,\vec{p})$ are real and
imaginary part of the retarded propagator, and
$\Sigma^R(E,\vec{p})$ is the retarded self-energy function.

The occurrence of this term can be understood when expressing
the action of the $j^\mu(x)$ current on the thermal groundstate of the
system only through the non-tildean fields. According to eq. (\ref{gse}),
\begin{eqnarray} \nonumber
\langle\mkern-4mu\langle 1| (-\i\gamma^0\widetilde{\psi}_y)\, \psi_x
&=&\langle\mkern-4mu\langle 1| \overline{\psi}_y\, \psi_x \\
\langle\mkern-4mu\langle 1| \overline{\psi}_x \,
  (-\i\widetilde{\overline{\psi}}_y\gamma^0)
&=&-\langle\mkern-4mu\langle 1| \psi_y\, \overline{\psi}_x
 = \langle\mkern-4mu\langle 1|
\left( \overline{\psi}_x \psi_y\, - \left\{ \overline{\psi}_x, \psi_y \right\}
\right)
\;.\end{eqnarray}
The calculation of the thermal groundstate expectation value of $j^\mu(x)$
current is rather tedious and requires a careful consideration of the
step functions under the integral. It is real and equal to
\begin{eqnarray} \nonumber
\left\langle j^\mu(x) \right\rangle &=&
\delta^{\mu 0}\;\int\!\!dE\,\int\!\!\frac{d^3\vec{p}}{(2\pi)^3}\,n(E)\, \\
\label{cc1}
&&
\mbox{Tr}\left[ \left(\gamma^0 -
  \frac{\partial \mbox{Re}\Sigma(E,\vec{p})}{\partial E}\right)\,
  {\cal A}(E,\vec{p})
+ \frac{\partial \mbox{Im}\Sigma^R(E,\vec{p})}{\pi\,\partial E}\,
  \mbox{Re}G(E,\vec{p})\right]
\;.\end{eqnarray}
Actually this is identical to the thermodynamical particle density
for particles with a nontrivial spectral function \cite{f95}, i.e.,
$\left\langle j^\mu(x)\right\rangle =
\delta^{\mu 0}\; \partial \Omega/\partial \mu $
where $\Omega$ is the thermodynamical potential of the effective field theory
and $\mu$ is the chemical potential.
Hence, all the requirements for the physical current operator are fulfilled by
our construction.

Finally we discuss the question where the above derivation might be useful.
To this end, we assume that our model lagrangian describes only
a part of an interacting system, where gauge bosons
(photons) are coupled to fermions with a nontrivial spectral function.
An excellent example for such a system might be a strongly
interacting plasma of quarks (plus other electrically neutral particles).

In such a system we want to calculate the radiation rate of thermal
photons, i.e., the ``glow'' of the plasma \cite{qh95gam}.
In principle, this has to be achieved
by considering the retarded physical current-current correlation function
\begin{equation}
\Pi^R_{\mu\nu}(x-y) = \Theta(x^0-y^0)\,
\left\langle j_\mu(x)j_\nu(y) \vphantom{\int}\right\rangle
\;.\end{equation}
If we consider this in momentum space,
and fix the coordinate system such that the 3-axis is aligned with
the photon emission direction, the emission rate of real photons
with energy (=momentum) $k$ is \cite{kap}
\begin{eqnarray} \label{ratea}
R &=&  2\frac{n_B(k,T)}{8 \pi^3}\,
   \mbox{Im}\left(\Pi^R_{11}+\Pi^R_{22}\right)\\
\label{rateb}
&=&2\frac{n_B(k,T)}{8 \pi^3}\,
   \mbox{Im}\left(-\Pi^{R\mu}_\mu\right)
   \;\;\mbox{ only if}\;\; k^\mu\,\Pi_{\mu\nu}(k)=0
\;.\end{eqnarray}
In principle, this calculation would be possible given our conserved
current (\ref{conc}) -- but the result would involve
the irreducible vertex function to a given order, and hence might
be difficult to obtain.

One may therefore ask, in which cases it is sufficient to
calculate the retarded one-loop fermion polarization tensor for
our nonlocal effective field theory,
\begin{equation} \label{eqimpi}
\Pi^{[1]R}_{\mu\nu}(k) = -\i \int\!\!\frac{d^4p}{(2\pi)^4}\;
  \mbox{Tr}\left[ g\,\gamma_\mu \, S^R(p+k)
    g\,\gamma_\nu \, S^A(p_0,\vec{p})\right]
\;,\end{equation}
which differs from the correlation function in two aspects:
It does not have the proper vertex factor (which is of higher loop
order), and it also misses the $\delta^{\mu0}$-corrections of (\ref{conc}).

Clearly this is the case if and only if the retarded self energy function
$\Sigma^R(E,\vec{p})$ is
a function of the energy $E$ alone (and {\em independent\/} of the
three-momentum $\vec{p}$ of the particle). The other matrix elements of
$\widehat{\Sigma}$ are obtained from the retarded self energy by
a transformation depending only on this energy parameter
\cite{h94rep}, and hence the vertex correction according to
eq. (\ref{exvex}) $\Lambda^\mu$ is also proportional to $\delta^{\mu0}$.

Independence of $\Sigma^R$ from the momentum might be good approximation
especially for fermions coupled to massless bosons \cite{h94spec},
and it follows that
\begin{equation}
\frac{\partial \Sigma^R(E,\vec{p})}{\partial \vec{p}} \approx 0
 \Rightarrow \Pi^{[1]R}_{ij}(k) = \Pi^{R}_{ij}(k),\;\;\;i,j=1,2,3
\;;\end{equation}
whereas $\Pi^{[1]R}_{0\mu}(k) \not= \Pi^{R}_{0\mu}(k)$. Hence,
for systems approximated by a momentum-independent retarded
self energy function, one {\em may} calculate the photon radiation rate
from the effective one-loop polarization tensor by use of eq.
(\ref{ratea}), but {\em not\/} by use of eq. (\ref{rateb}) \cite{qh95gam}.

With the present paper we reached two goals. First of all, we
have established a clean path towards current-conserving
(and thus gauge invariant) calculations in generalized free field
theory. As we have shown, this is especially relevant for
the determination of emission rates of gauge bosons (photons) from a hot
plasma. Secondly, we have constructed a conserved {\em physical\/}
current operator in the framework of thermo field dynamics,
which is thermodynamically correct. To our knowledge, this has
not been achieved in other formulations of thermal field theory,
we are currently exploring this connection in more detail.
\subsubsection*{Acknowledgement}
We acknowledge stimulating discussions with B.Friman, E.Quack,
L.Razumov and G.Wolf.

\end{document}